\documentclass[
10pt,twocolumn,notitlepage,oneside]{revtex4}
\usepackage{rotating}
\usepackage{amsfonts,longtable}

\begin{document}

\title{Galaxies with Unusually High Abundances of Molecular Hydrogen}
\author{\firstname{A.~V.}~\surname{Kasparova}}
\email[]{akasparova@yandex.ru}
\affiliation{Sternberg Astronomical Institute. }
\author{\firstname{A.~V.}~\surname{Zasov}}
\email[]{zasov@sai.msu.ru}
\affiliation{Sternberg Astronomical Institute.}
%
\begin{abstract}
A sample of 66 galaxies from the catalog of Bettoni et al. (CISM)
with anomalously high molecular-to-atomic hydrogen mass ratios
$(M_{mol}/M_{HI}>2)$ is considered. The sample galaxies do not
differ systematically from other galaxies in the catalog with the
same morphological types, in terms of their photometric parameters,
rotational velocities, dust contents, or the total mass of gas in
comparison with galaxies of similar linear sizes and disk angular
momentum. This suggests that the overabundance of $H_2$ is due to
transition of $HI \to H_2$. Galaxies with bars and active nuclei are
found more frequently among galaxies which have $M_{mol}$ estimates
in CISM. In a small fraction of galaxies, high $M_{mol}/M_{HI}$
ratios are caused by the overestimation of $M_{mol}$ due to a low
conversion factor for the translation of {\it CO}-line intensities
into the number of $H_2$ molecules along the line of sight. It is
argued that the "molecularization" of the bulk of the gas mass could
be due 1) to the concentration of gas in the inner regions of the
galactic disks, resulting to a high gas pressure and 2) to
relatively low star-formation rate per unit mass of molecular gas
which indeed takes place in galaxies with high $M_{mol}/M_{HI}$
ratios.

Astronomy Reports, 2006, Vol. 50, No. 8, p. 626--637.

\end{abstract}
\maketitle
\section{INTRODUCTION}

The densest and coolest component of the interstellar medium~---
molecular gas~--- is observed in all types of disk galaxies. In many
cases, a mass of molecular gas inside galactic disks is comparable
to a mass of neutral hydrogen ($HI$). The amount and spatial density
of molecular gas reflects the specific properties of the evolution
of the interstellar medium in galactic disks. The most massive
gaseous formations~--- giant molecular clouds, which affect the
dynamical evolution of the disk,--- are associated with molecular
gas. However, more importantly, the mass and density of the
molecular gas determine the star-formation rate and star-formation
efficiency in galactic disks.

Despite the crucial role that molecular gas plays in galaxies, the
factors determining its abundance relative to neutral gas remain
poorly known. Most information on interstellar molecular gas has
been obtained from measurements of {\it CO}-line intensities~---
{\it CO} has a very low excitation temperature. The conversion of
the {\it CO}-line luminosity into the total mass of gas is a
separate problem. For galaxies with normal heavy-element abundances
the conversion factor connecting the {\it CO}-line intensity with
the number of $H_2$ molecules along the line of sight, ${\it X} =
N(H_2)/I_{CO}$, is usually set equal to $(2-3)\cdot 10^{20} \quad
mol/(K \cdot km/s)$, however, the universality of this parameter
remains a topic of discussion. Adopting a constant ${\it X}$ value
is, of course, a simplification, especially when galaxies with
strongly differing mass, gas density, or star-formation intensity
are compared. Various methods for measuring ${\it X}$ and the
dependence of ${\it X}$ on the galactic luminosity show appreciable
scatter, together with a systematic increase in ${\it X}$ with
decreasing luminosity (mass) of the galaxy and decreasing
heavy-element abundance (\cite{boselli}, \cite{arimoto}). The
conversion factors for dwarf galaxies and giant luminous galaxies
may differ by an order of magnitude. We also expect ${\it X}$ to be
smaller for dense circumnuclear disks, where the molecular gas seems
to be contained in both clouds and the diffuse intercloud medium
(\cite{downes}, \cite{rosolovsky}).

Various authors (\cite{elmegreen89}, \cite{elmegreen93},
\cite{wong}) have analyzed the conditions for molecular-to-atomic,
$M_{mol}\to M_{HI}$, transformations and vice-versa (see also
references in these papers). The rate at which molecules are formed
depends on both the density (or pressure) of a gas and on the
density of {\it UV} flux, which destroys molecules. According to
Elmegreen (\cite{elmegreen93}), the $M_{mol}/M_{HI}$ ratio is
proportional to $P^{2.2}/j$, where $P$~is the gas pressure and
~$j$~is the {\it UV} flux density. The gas compression in shocks can
also play an important role in the "molecularization" of
hydrogen~\cite{kee}.

Since the gas pressure depends both on its surface density and on
the volume density of the stellar disk, it decreases rapidly with
galactocentric distance. Indeed, the azimuthally averaged density
ratio $M_{mol}/M_{HI}$ follows the pressure in such a way that the
contribution of $M_{mol}$ to the total mass of gas becomes small at
a periphery of the galaxy~\cite{wong}.

Available estimates show that the integrated ratio $M_{mol}/M_{HI}$
is appreciably lower than unity for most spiral galaxies. According
to Casoli et al.~\cite{casoli}, $M_{mol}/M_{HI}$ is, on the average,
close to $0.2$. Boselli et al.~\cite{boselli} found the average
molecular-to-atomic gas mass ratio to be $M_{mol}/M_{HI} = 0.14$ for
a sample of 266 galaxies. However, some spiral galaxies are known to
have anomalously high $M_{mol}/M_{HI}$ values, exceeding two or even
three (an order of magnitude higher than the average value). Our aim
here is to analyze the specific features of such galaxies with a
predominance of molecular gas and possible factors that could lead
to such high $M_{mol}/M_{HI}$ estimates.

As our initial data, we used "A new catalogue of ISM content of
normal galaxies" (CISM) of Bettoni et~al. \cite{bettoni}. This
catalog contains interstellar-medium data for almost two thousand
"normal" galaxies, more than 300 of which have estimates for their
molecular-gas mass. Normal galaxies are considered to be those that
display no morphological distorsion (such as tidal tails or bridges)
or strong perturbations of their circular velocities, although
several galaxies among those included in the catalog can,
nevertheless, be considered to be interacting galaxies (see below).
The molecular-gas mass estimate is based on the adopted value of
${\it X} = 2.3\cdot 10^{20} \quad mol/K \cdot km/s$, taken to be the
same for all galaxies.

\section{MAIN PROPERTIES OF GALAXIES WITH A PREDOMINANCE OF $H_2$}

\subsection{The Galaxy Sample}
We selected a total of 66 objects with anomalously high ratios
$M_{mol}/M_{HI} > 2$ from the CISM catalog of Bettoni
et~al.~\cite{bettoni}. Forty-five of these galaxies have
$M_{mol}/M_{HI} > 3$. Table~\ref{tab1} gives the main properties of
these sample galaxies. We adopted data on the interstellar medium
from CISM; photometric and kinematic properties from the HYPERLEDA
\cite{hyperleda} database, and the photometric scale in accordance
with the catalog of Baggett et~al.~\cite{bagg}. The columns of the
table give: a running number for the galaxy, its name, morphological
type (as a number), heliocentric distance (in Mpc), radial
photomeric scale length (in kpc), {\it B}-band luminosity (in solar
luminosities), photometric diameter  $D_{25}$ (in kpc), logarithm of
the dust mass (in solar units), logarithm of the $HI$ mass (in solar
units), logarithm of the mass of molecular gas (in solar units),
logarithm of the rotational velocity derived from the widths of $HI$
(in km/s), and corrected {\it U-B} and~{\it B-V} color indices. The
last column ("Remarks") shows the presence of a bar (B), membership
in the Virgo or Coma clusters (V or C), and signs of interaction (VV
number according to Vorontsov-Velyaminov).

\begin{figure*}

\includegraphics[scale=0.9]{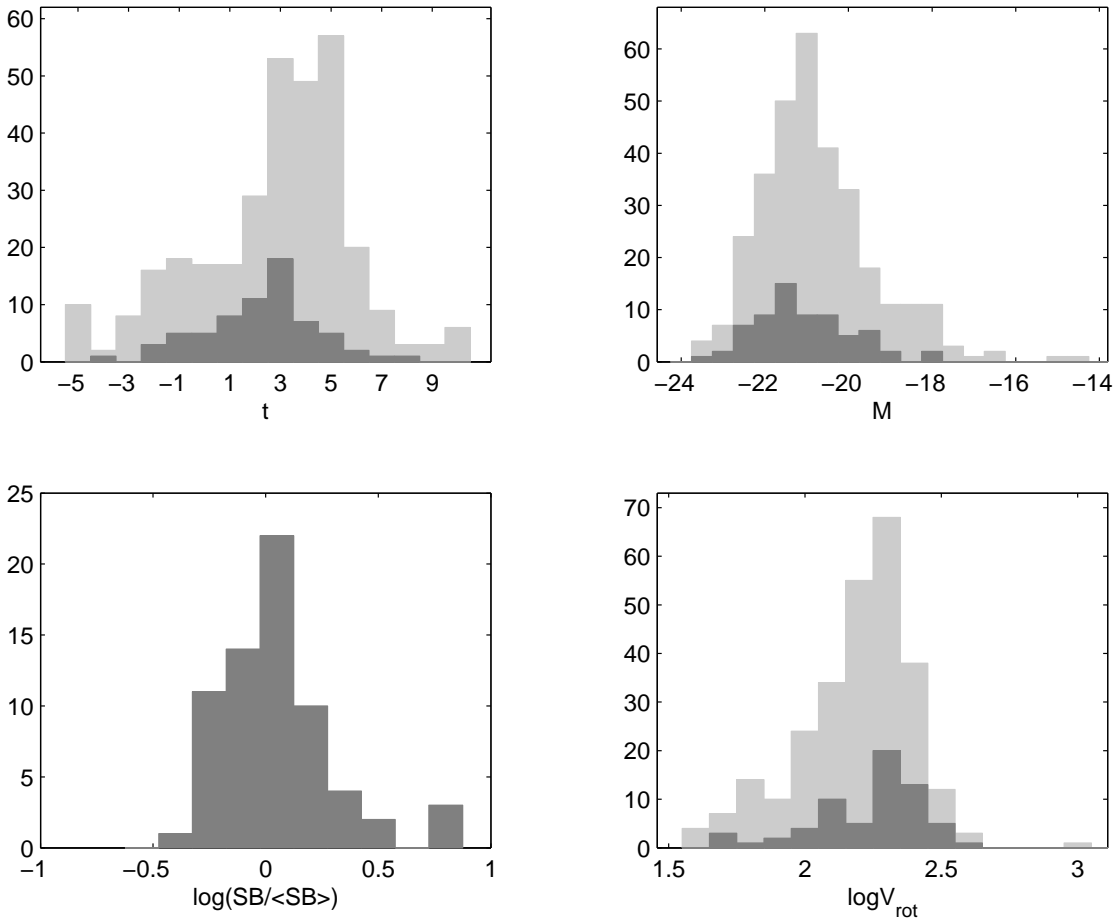}
\caption{Distributions of the galaxies in morphological type $t$,
absolute {\it B}-band magnitude $M$, logarithm of the surface
density normalized by the mean surface density for the corresponding
morphological type, and the logarithm of the rotational velocity,
$V_{rot}$. The light gray histograms show the distributions for all
galaxies of catalog \cite{bettoni} with estimated $M_{mol}$ values,
while the dark gray histograms are for galaxies in our sample, that
is a predominance of molecular gas.} \label{fig1}
\end{figure*}

Figure~\ref{fig1} compares the galaxies in our sample with the CISM
galaxies with known molecular-gas contents (below, we will consider
only this part of all CISM catalog) in terms of their morphological
types, surface brightnesses $SB$ (within the isophotal diameter
$D_{25}$) relative to the average surface brightness
$\left<SB\right>$ for CISM galaxies of the same type, absolute
B-band magnitude $M$, and rotational velocity $V_{rot}$ derived from
the $HI$ line widths. These distributions show that galaxies with
high $M_{mol}/M_{HI}$ ratios are mostly {\it Sa~---~Sbc} galaxies
($t = 1~-~4$), with virtually no late-type galaxies among them. Note
that allowing for the luminosity dependence of the conversion factor
does not affect the conclusion that galaxies with late morphological
types have lower $M_{mol}/M_{HI}$ ratios \cite{boker}.

The galaxies of our sample are essentially indistinguishable from
the other CISM galaxies in terms of the brightness of their stellar
disks and their luminosities, although the mean value for $M$ is
shifted somewhat toward higher luminosities. The rotational
velocities $V_{rot}$ for the sample galaxies and for all the CISM
galaxies with available $M_{mol}$ estimates have more or less the
same distributions, ranging from $50$ to $400$ $km/s$.
Table~\ref{tab2} gives the numerical results of these comparisons:
the mean values and standard deviations of the quantities compared,
and the probability that the difference between the means of the two
samples is of random nature. We conclude that the objects in our
sample do not differ strongly from the other galaxies in terms of
the parameters described above.

\begin{table}
 \caption{Mean values
of some parameters for galaxies of the CISM catalog (sample 1) and
for sample galaxies with $M_{mol}/M_{HI}>2$ (sample 2), with the
standard deviations (STD) and the statistical probability of the
absence of differences between the two samples (p).}

\bigskip
\begin{tabular}{cccc}
\hline \hline
Sample & Mean & STD & p, \% \\
\hline \hline
\multicolumn{4}{c}{Morphological type} \\
\hline
1 & 2.80 & 3.06 & $95.00$ \\
 2 & 2.24 & 2.24 &\\
\hline
\multicolumn{4}{c}{Magnitude} \\
\hline
1 & -20.75 & 1.37 & $99.00$ \\
2 & -21.06 & 1.16 &\\
\hline
\multicolumn{4}{c}{$logV_{rot}$} \\
\hline
1 & 2.19 & 0.21 & $95.00$ \\
2 & 2.23 & 0.20 &\\
\hline
\multicolumn{4}{c}{$log(M_{dust}/M_{gas})$} \\
\hline
1 & -3.38 & 0.39 & $99.95$ \\
2 & -3.21 & 0.36 &\\
\hline
\end{tabular}
\label{tab2}
\end{table}

Note that galaxies with optical bars make up a significant fraction
of galaxies with measured molecular-gas masses~(Table~\ref{tab3}).
Whereas barred galaxies account for about one-third of all the CISM
galaxies, they comprise one-half of all galaxies with available
molecular-gas mass estimates (which evidently bias to objects with
high $H_2$ abundances). The latter is also true for the galaxies in
our sample with~$M_{mol}/M_{HI} > 2$. It follows from
Table~\ref{tab3} that this sample is characterized by high frequency
of occurrences of bars and of nuclear activity (the latter was taken
from the catalog~\cite{veron}). This may be indirectly associated
with the concentration of a gas toward galactic centers. The
existence of a possible correlation between the abundance of
molecular gas and the level of nuclear activity was earlier pointed
out in~\cite{evans}.

Thus, galaxies with high $M_{mol}/M_{HI}$ ratios do not differ
significantly from other galaxies in their photometric properties,
morphological properties, or rotational velocities. However, the
question remains open as to what extent the abundance of atomic and
molecular gas in such galaxies is really anomalous.

\begin{table}
 \caption{Percentage
of barred galaxies and galaxies with active nuclei (according to
\cite{veron}) for (1) all objects of CISM \cite{bettoni}, (2)
galaxies of CISM \cite{bettoni} with molecular-hydrogen mass
estimates, and (3) the galaxies of our sample with high relative
abundances of molecular gas. }

\bigskip
\begin{tabular}{cccc}
\hline \hline
Sample & Bar & Active nucleus & N  \\
\hline
1 & $35.3 \pm 0.8$ & $7.7 \pm 0.2$ & 1916 \\
2 & $53.3 \pm 3.0$ & $15.8 \pm 0.9$& 317 \\
3 & $56.7 \pm 6.9$ & $22.4 \pm 2.7$& 66 \\
\hline \hline
\end{tabular}
\label{tab3}
\end{table}

\subsection{$HI$ and $H_2$ Abundances in the Sample Galaxies}
\begin{figure*}
\includegraphics[scale=0.9]{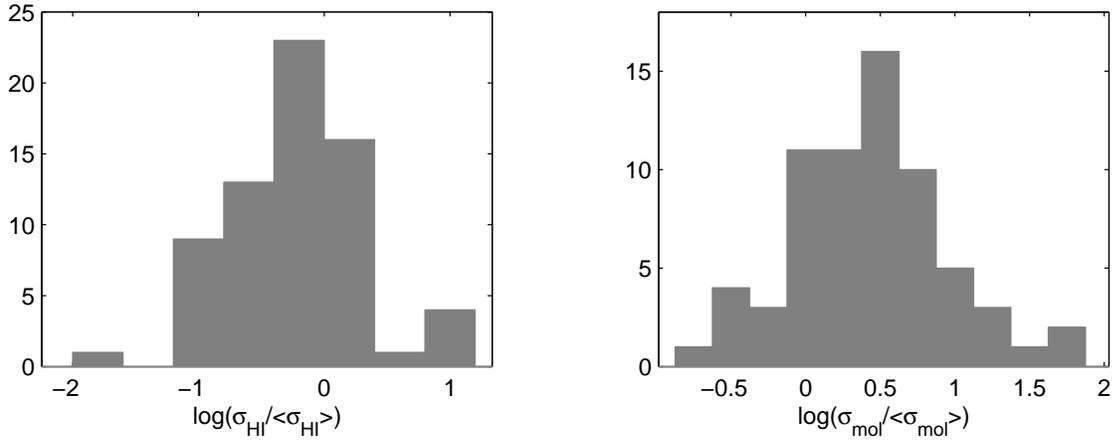}
\caption{Histograms of $HI$ and the molecular-gas surface density
normalized to the mean values for the corresponding morphological
types for the galaxies of the sample studied.} \label{fig2}
\end{figure*}

A high $M_{mol}/M_{HI}$ ratio could be due to either a high
abundance of molecular gas or a reduced mass $HI$ in the galaxy. Let
us now compare the galaxies in our sample with $M_{mol}/M_{HI} > 2$
and all the CISM galaxies of the corresponding morphological types.
The left-hand panel in Fig.~\ref{fig2} shows the distribution of the
sample galaxies in $lg(\sigma_{mol}/\left<\sigma_{mol}\right>)$,
where $\sigma_{mol}=M_{mol}/D^2_{25}$ is the surface density of the
molecular gas within the optical radius, and
$\left<\sigma_{mol}\right>$ is the mean $\sigma_{mol}$ for all
galaxies of the same morphological type with known masses $M_{mol}$.
The right-hand panel in Fig.~\ref{fig2} shows a similar comparison
of the surface density of neutral hydrogen, $\sigma_{HI}$. The mean
$\sigma_{mol}$ for the sample galaxies is higher by $ \triangle
\lg(\sigma_{mol} /\left<\sigma_{mol}\right>) \approx 0.5$ compared
to the average value for all galaxies of the corresponding type. The
$HI$ abundance in the sample galaxies is characterized by an
asymmetric shape of the distribution of $\sigma_{HI}$ (note the more
gently sloping "wing" for galaxies with strong deficiencies of
neutral hydrogen), although the mean $\sigma_{HI}$ does not differ
strongly from the value $\left<\sigma_{HI}\right>$ for all galaxies
of the same morphological type. If we select galaxies with the
lowest $HI$ abundances, with surface densities $\sigma_{HI}$ less
than half the mean value, we will find that their mean
$\lg(\sigma_{mol}/\left<\sigma_{mol}\right>)$ is also higher than
zero, indicating that, in this case too, we are dealing with an
excess of molecular gas. Thus, galaxies with high $M_{mol}/M_{HI}$
ratios really possess high $H_2$ abundances, rather than having lost
their $HI$ as a result of, e.g., its being swept out from the galaxy
as a more tenuous component of the interstellar medium.

\subsection{Total Gas Mass in the Disks of the Sample Galaxies}
We now compare the total gas contents of the galaxies using the
relation between the gas mass and the specific angular momentum of
the disk, which is characterized by the product of the rotational
velocity $V_{rot}$ and radius $R$ of the disk. As the disk radius,
we used either the photometric (isophotal) radius $R_{25}$ or the
photometric radial disk scale $R_0$ (in linear units). The relation
between the mass of gas and $V_{rot}R$ is obeyed by most of the
Sb~---~Ir galaxies with various brightnesses, both single objects
and galaxies with close companions~\cite{zasov}, \cite{karach} (see
also references therein). Although this relation has a physical
basis, and follows from the fact that the gas surface densities in
most of the galaxies are (on average) close to the threshold density
for local Jeans instability, or proportional to this density, here
we treat it as a purely empirical relation.

\begin{figure*}
\includegraphics[scale=0.8]{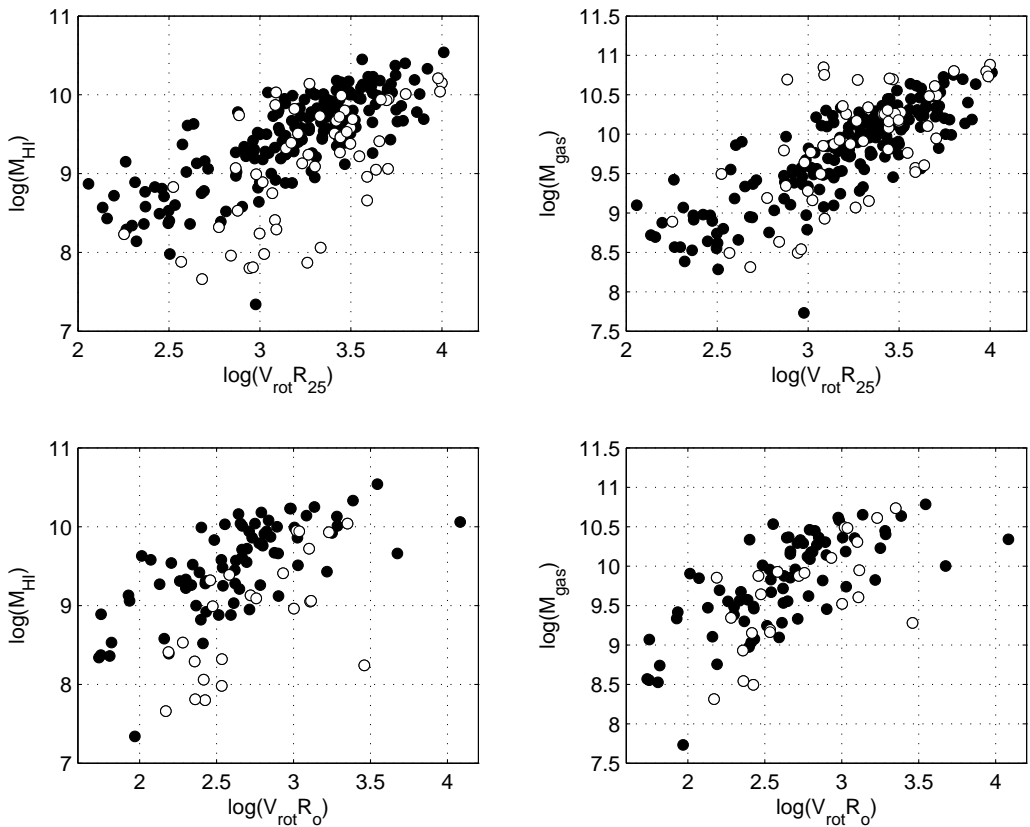}
\caption{Mass of atomic gas (left) and total mass of gas (right) as
a function of specific angular momentum defined as $V_{rot}R_{25}$
(top) or $V_{rot}R_0$ (bottom) for galaxies of the sample considered
(open circles) and for all galaxies of catalog \cite{bettoni} for
which the corresponding estimates are available (filled circles).
$R_{25}$ and $R_0$ are photometric radius and a radial scale length
of a disk.} \label{fig3}
\end{figure*}

Figure~\ref{fig3} (left, top and bottom) shows the dependence of the
mass of atomic gas $M_{HI}$ on the specific angular momentum. All
set of CISM galaxies and galaxies from our sample are shown by
filled and open circles, respectively. Most of the sample galaxies
display $HI$ deficits, which are especially strong among galaxies
with relatively low angular momenta. The situation changes radically
if instead of $M_{HI}$ we plot the total mass of atomic and
molecular gas (including helium) $M_{gas}=1.4(M_{HI}+M_{mol})$ on
the vertical axis (Fig.~\ref{fig3}, top right). In this plot, the
sample galaxies come closer to the universal relation. Only a few
systems fall away from the overall relation because they possess
unusually high masses of gas. Their $M_{mol}$ values are apparently
overestimated (see the next section). Thus, the total gas masses for
most of the galaxies with anomalously high molecular-gas abundances
are consistent with the angular momenta of the galaxies. Hence, the
excess $H_2$ in such galaxies is not neither due to excess
"portions" of molecular gas that have come to the galaxy disk, e.g.,
via the accretion of cold clouds, nor due to overestimates of the
$H_2$ abundances, but instead it is caused by the transition of most
of the interstellar hydrogen into the molecular state.

This conclusion remains unchanged if the radial photometric disk
scale $R_0$ is used instead of the optical radius (lower panel of
Fig.~\ref{fig3}). Note that unlike the isophotal radius, $R_0$ is
not sensitive to differences between the surface brightnesses of the
disks of the galaxies compared.

\section{POSSIBLE ORIGINS OF HIGH $M_{mol}/M_{HI}$ RATIOS}
Let us now list the main reasons that can lead to high estimates of
the relative mass of molecular gas.
\begin{enumerate}
    \item The actual conversion factors for some galaxies
    may be much lower than the adopted value.
    \item This is the effect of the galaxy environment (accretion of large masses of molecular gas onto the disk, or the displacement inward of very cold, difficult to detect, molecular gas from the disk periphery due to interactions between
    galaxies~\cite{pfenniger}).
    \item An anomalously high content of dust, which serves as a catalyst of the formation of molecules and prevents the dissociation of molecules by the {\it UV} radiation of stars.
    \item Longer lifetimes of the gas in molecular form compared to most galaxies (e.g., a low star-formation rate per unit mass of molecular gas).
    \item Higher gas pressure in regions where it is mainly concentrated, which stimulates the transition of $HI$ into $H_2$.
\end{enumerate}
Bellow we analyze each of the above possibilities one-by-one as
applied to our sample of galaxies with high $M_{mol}/M_{HI}$ ratios.

\subsection{Conversion Factor}
The conversion factor ${\it X}$ may be higher for galaxies with
relatively high heavy-element abundances \cite{boselli}. Metallicity
of gas is usually associated with a high total luminosity $L$ of a
galaxy. Since the galaxies in our sample are not characterized by
very high luminosities (Fig.~\ref{fig1}), this explanation is not
appropriate for most of our galaxies.

\begin{figure*}
\includegraphics[scale=0.8]{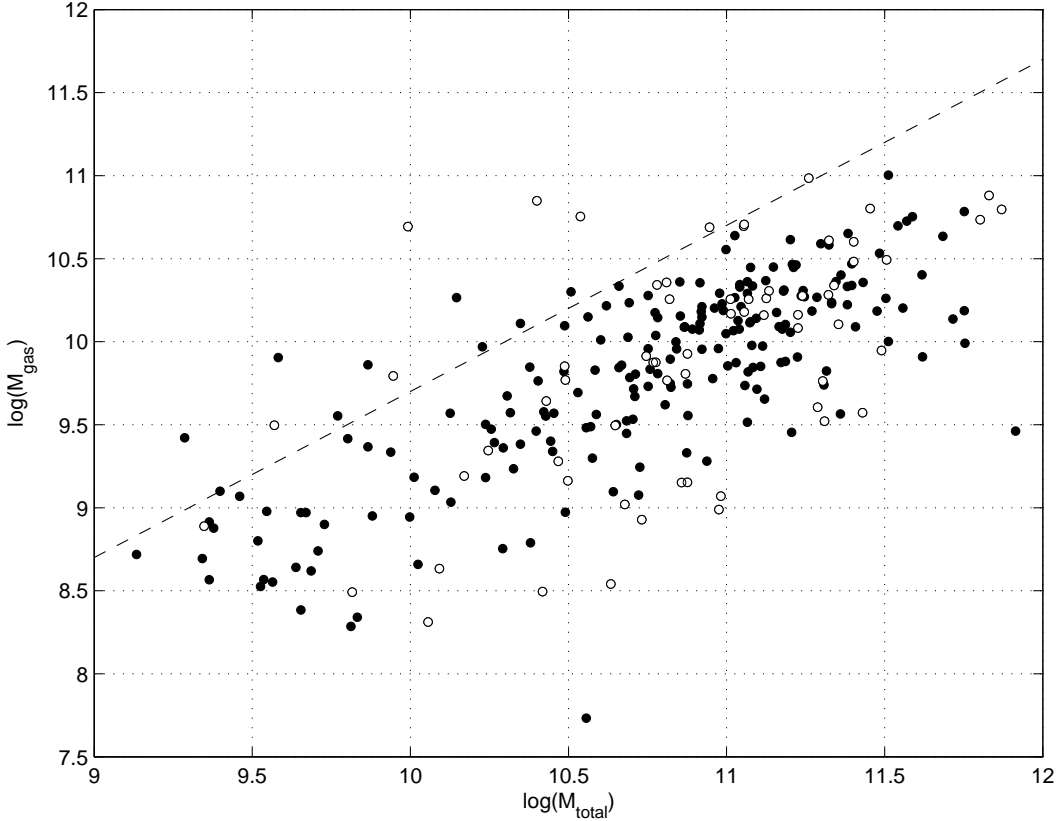}
\caption{Estimated total gas mass (including helium) $M_{gas}$ as a
function of the total (indicative) mass of the galaxy $M_{tot}$,
within the photometric radius $R_{25}$. Filled and open circles
correspond to CISM \cite{bettoni} galaxies and to galaxies in our
sample, respectively. The dashed line corresponds to
$M_{gas}=M_{tot}/2$. The mass $M_{mol}$ is likely substantially
overestimated for galaxies located above this line.} \label{fig4}
\end{figure*}

However, some of our sample galaxies may indeed have anomalously low
${\it X}$ values. Grounds for this possibility are provided by the
fact that the estimated total gas mass $M_{gas}$ (including helium)
derived for the adopted ${\it X}$ is unrealistically high~---
comparable to the total mass of the galaxy $M_{tot}$ within its
optical boundaries. It is obvious that $M_{tot}$ must exceed the
total mass of the disk (not to mention the mass of its gaseous
component) if to take into account the masses of the dark halo and
bulge. Figure~\ref{fig4} compares the total gas mass with indicative
mass $M_{tot} = V_{rot}^2D_{25}/2G$ (in solar units), which is
approximately equal to a total mass of a galaxy within $D_{25}$, for
all the CISM galaxies with available estimates of $V_{rot}$ and
$D_{25}$. For galaxies located above the slanting line,
$M_{gas}>M_{tot}/2$, and the gas masses are very likely to be
overestimated. These galaxies make up a small fraction of all the
galaxies considered, but about half of these objects belong to our
sample of galaxies with a predominance of molecular gas (open
circles). The same galaxies also exhibit a gas excess in the
diagrams shown in Fig.~\ref{fig3}, supporting the hypothesis that
the conversion factors for these objects may have been strongly
overestimated. Note that the exclusion of these galaxies does not
alter the conclusions drawn above.

Lower conversion factors may be associated not only with high
metallicity of a gas, but also with the specific physical state of
the interstellar medium. For example, the conversion factor should
be low if molecular gas is concentrated not in virialized clouds, as
it is usually assumed, but in a less dense diffuse
medium~\cite{downes}, which is quite possible in the case of the
densest circumnuclear disks, where both clouds and the more tenuous
intercloud gas are molecular. The example of {\it M64}, which dense
molecular circumnuclear disk is observed in {\it${}^{12}CO$}
and~{\it ${}^{13}CO$} lines, shows that diffuse molecular gas not
only exists, but may even be responsible for the bulk of the {\it
CO}-line luminosity, and the use of the standard conversion factor
yields appreciably an overestimated mass of molecular gas
\cite{rosolovsky}.

\subsection{Effect of the Environment}
The effect of the environment is not the main factor determining the
high $M_{mol}/M_{HI}$ ratios for the galaxies in our sample. First,
the sample contains only a few strongly interacting systems or
galaxies in clusters (Table~\ref{tab1}). Second, with few
exceptions, dense environments around galaxies are not associated
with a sharp increase in $M_{mol}/M_{HI}$. According to de Mello
et~al.~\cite {demello}], single galaxies and galaxies located in
more crowded environments differ only slightly in terms of
$M_{mol}/M_{HI}$, and the mean $M_{mol}/M_{HI}$ ratio for such
galaxies remains less than unity. They report
$\left<\lg(M_{mol}/M_{HI})\right>=-0.69\pm0.59$ and~$-0.51\pm0.46$
for the the means and standard deviations for the high-density
sample and the control sample of isolated galaxies, respectively.

\subsection{High Dust Content}
\begin{figure*}
\includegraphics[scale=0.8]{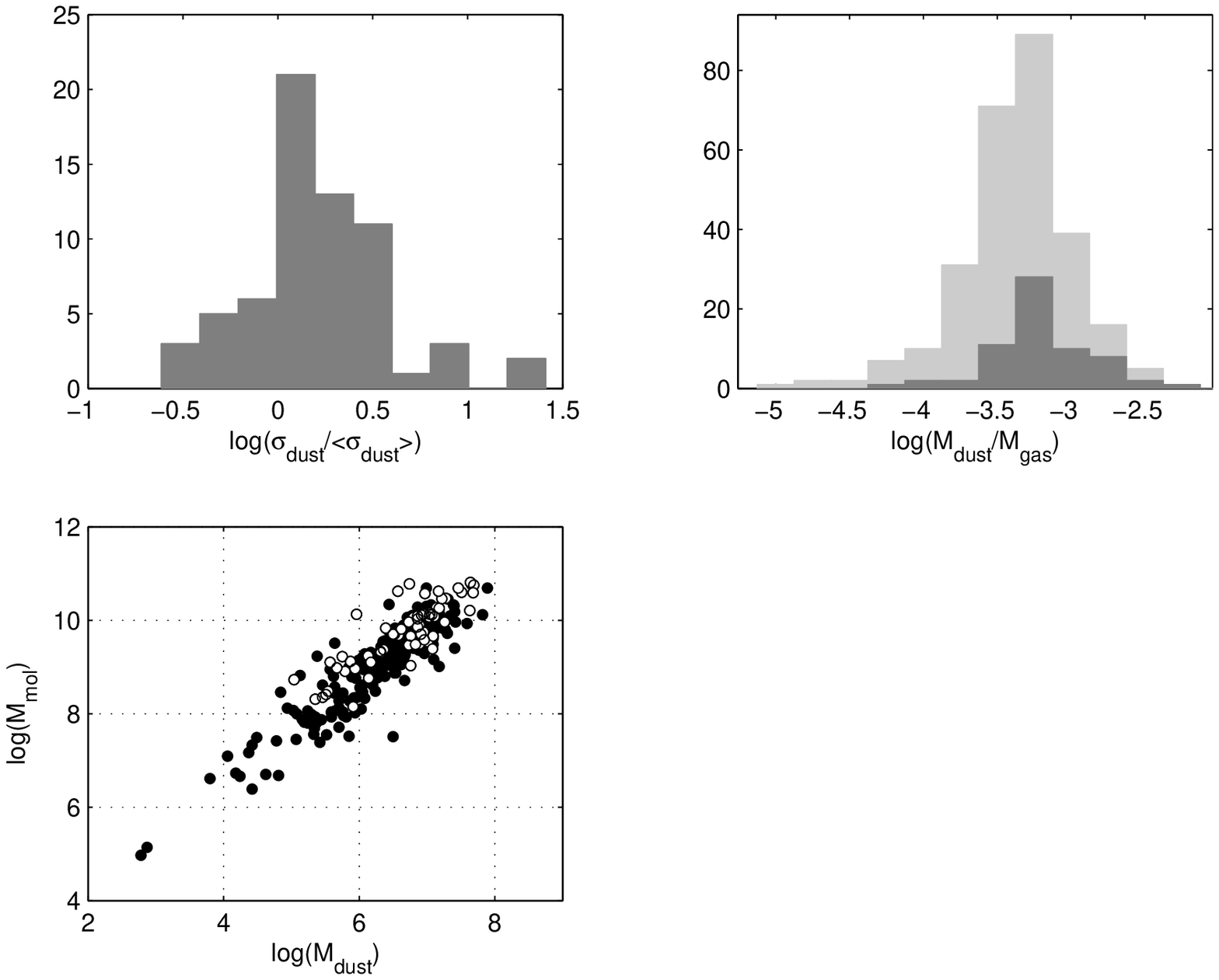}
\caption{Dust content in galaxies. Top left: a histogram of the
logarithm of the ratio of the surface density of dust to the mean
value for the given morphological type according to the CISM catalog
\cite{bettoni}; top right: a histogram of the logarithm of the ratio
of the mass of dust to the mass of gas (dark gray is for our sample
of galaxies, light gray is for all galaxies of the CISM catalog with
available $M_{mol}$); and at the bottom: the dependence of the mass
of molecular gas on the mass of dust (filled circles are for all
CISM galaxies, while open circles are for galaxies in our sample).}
\label{fig5}
\end{figure*}

\begin{figure*}
\includegraphics[scale=0.8]{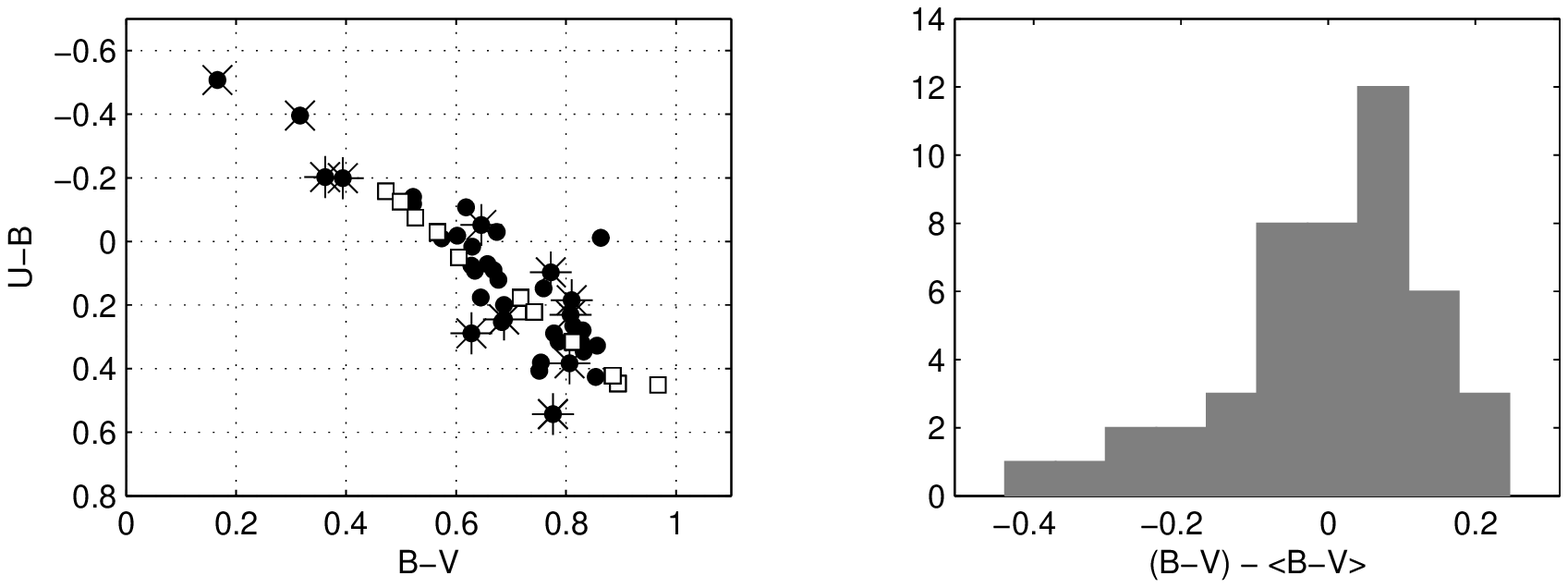}
\caption{Left: two-color diagram for the corrected color indices
$(U-B)$~---~$(B-V)$ (according to the HYPERLEDA database). Filled
circles show the galaxies in our sample and squares are for the mean
values for galaxies of various morphological types (according to
\cite{buta}). The X's denote interacting galaxies and the asterisks
are for galaxies that belong to clusters. Right: distribution of the
galaxies in our sample in the deviation of their $B-V$ color indices
from the mean values for the corresponding morphological type).}
\label{fig6}
\end{figure*}

Let us now compare the galaxies in terms of the mass of dust they
contain, estimated from the far infrared (FIR) radiation of
galaxies. Figure~\ref{fig5} (top left) compares the dust surface
density defined as $M_{dust}/D^2_{25}$ in our sample galaxies with
the mean dust density for all CISM galaxies of the given
morphological type. It is evident that galaxies with a predominance
of molecular gas are, indeed, characterized by a higher mean dust
content. However, this is probably due solely to the higher mass of
gas in these galaxies, since the distribution of $M_{dust}/M_{gas}$
for our galaxies differs little from that for all the CISM galaxies
(Fig.~\ref{fig5}, top right). Note that the $M_{dust}$ estimate used
here refers to so-called warm dust heated by stars, which emits
predominantly in the FIR ($40 \div 120$ $\mu m$). The mass of dust
at very low temperatures, which weakly radiates in this spectral
interval, cannot be taken into account rigorously. However, there
are no grounds to believe that galaxies with a high content of
molecular gas detected in {\it CO} are also characterized by
anomalous amounts of very cool dust. The location of the sample
galaxies on a two-color diagram likewise does not indicate high
color excesses (Fig.~\ref{fig6}). The squares on the diagram show
the mean color indices for galaxies of various morphological
types~\cite{buta}, which mark out the location of the normal color
sequence.

%

Observations show that the mass of dust in galaxies is most closely
related to the mass of molecular gas~\cite{devereux}.
Figure~\ref{fig5} shows the $M_{dust}$~---~$M_{mol}$ diagram for
CISM galaxies. The sample galaxies (open circles) lie on the common
relation (lower panel in Fig.~\ref{fig5}), confirming that these
objects have a "normal" content of dust per unit mass of molecular
gas (except for a few galaxies).

\subsection{Long Lifetime of Gas in the Molecular State}

Molecular gas gives birth to stars whose radiation, in turn, leads
to the dissociation of molecules. Therefore, the hypothesis that gas
spends a long time in the molecular state resulting in the large
$M_{mol}$ is equivalent to suggestion of a low star-formation rate
per unit mass of molecular gas. It is worth noting that most of the
sample galaxies with high $M_{mol}/M_{HI}$ are not characterized by
high star-formation rates, as is testified to by their location in
the two-color diagram (Fig.~\ref{fig6}). Two objects in the~"blue"
part of the diagram, marked by crosses, are interacting systems or
"mergers" ({\it MCG~5-32-20} and~{\it MCG~5-29-45}), which appear to
be undergoing bursts of star formation. Judging by their color
indices, most of the galaxies are characterized by moderate or low
star-formation rates. The only galaxy that falls away from the
overall dependence due to its high {\it (B-V)} is {\it NGC1482},
which is a peculiar lenticular galaxy undergoing a burst of star
formation in its central region, and is characterized by an extended
region of bright line emission and gas outflow (superwind)
\cite{veilleux}.

Star formation is a multistage process. If star formation proceeds
in a quasi-stationary mode, the mass of gas involved in each stage
of the process is proportional to the duration of this stage. In
this case, a large amount of molecular gas could be due to a
relatively slow formation of gravitationally unstable regions inside
molecular clouds (inefficient damping of turbulent motions?) or in
the diffuse molecular gas, which evolves on a shorter time scale
than the dense clouds.

\begin{figure*}
\includegraphics[scale=0.8]{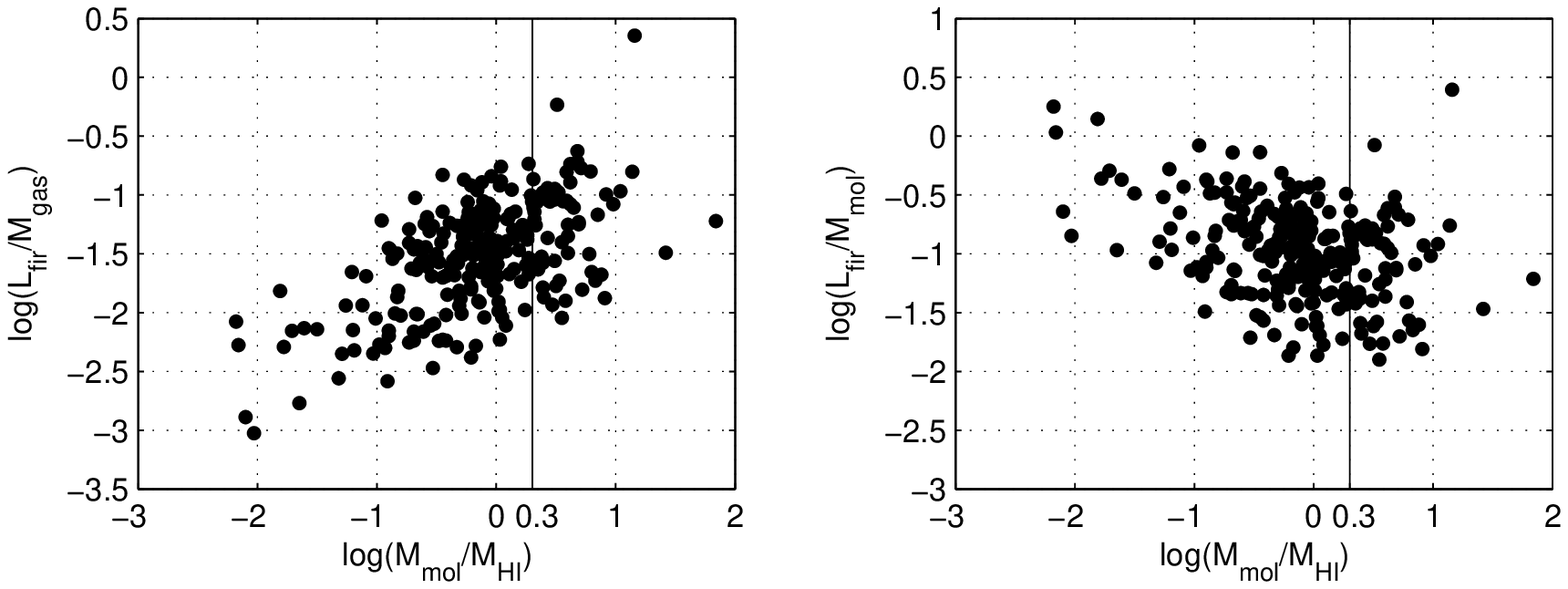}
\caption{Dependence of the star-formation efficiency defined as
$log(L_{FIR}/M_{gas})$  and as $log(L_{FIR}/M_{mol})$ (right) on the
relative content of molecular gas, $log(M_{mol}/M_{HI}$), for CISM
galaxies; the vertical line separates the sample galaxies with hight
fraction of $M_{mol}$.} \label{fig7}
\end{figure*}

The current star-formation rate is closely related to the infrared
emission by dust $L_{FIR}$, which absorbs the short-wavelength
radiation of stars. To compare the star-formation efficiencies of
the galaxies, the left and right panels of Fig.~\ref{fig7} show the
relative mass of gas for the CISM galaxies with $L_{FIR}/M_{gas}$
and $L_{FIR}/M_{mol}$, respectively. The first dependence
demonstrates an increase in the star-formation efficiency with an
increasing relative $H_2$ content: the interstellar gas is used up
faster in galaxies that are rich in molecular hydrogen than in those
with small amounts of $H_2$. However, this is true only for the
total mass of gas, which in most cases is close to the mass of $HI$.
However, as follows from the right-hand plot, the larger molecular
fraction of gas, the lower, on the average, the star-formation rate
per unit mass of molecular gas. Consequently, the intensity of the
radiation is lower leading to molecular dissociation. This means
that, on average, in galaxies dominated by $H_2$, the gas should
spend more time in the molecular state before it becomes involved in
the process of star-formation and dissociation.

On both diagrams, two galaxies deviating strongly from the overall
dependence show up conspicuously in the top right corner are {\it
IC860} and {\it NGC4418}. The anomalously high intensity of their
infrared radiation is likely associated with bursts of star
formation in these objects \cite{kazes}, \cite{imanishi}.

\subsection{Excess Pressure of the Interstellar Gas}

The pressure of gas is determined mainly by two parameters~--- the
gas surface brightness and the volume density of the disk, which is
in good agreement with observations \cite{wong}, \cite{blitz},
\cite{heyer}. A high gas pressure appears to be a crucial factor
leading to the transformation of neutral (into molecular) hydrogen
\cite{elmegreen93}.

A high pressure (density) of gas can be due both to the existence of
local regions of compression (high contrast spiral arms, powerful
star-forming regions associated with superclouds) and to a higher
density of the stellar disk in the regions containing most of the
gas. {\it M51} is an example of a galaxy with a high relative
content of molecular gas in its powerful spiral arms \cite{aalto}.
Giant gaseous complexes are concentrated in the spiral arms, and the
gas velocity fields indicate compression of the gas at the front of
the shock associated with them. In the absence of strong
compressional waves, the gas would first be transformed into
molecular form in regions with higher surface density and smaller
thicknesses of the gaseous disk. Therefore, the highest gas
pressures are expected where gas is concentrated in the inner region
of the galaxy. Indeed, about half the gas-rich spiral galaxies
(although these do not necessarily show a predominance of $H_2$ over
$HI$) with available high resolution data on the radial distribution
of the {\it CO} intensity demonstrate a significant growth of the
density of molecular gas toward the center in the inner-disk region
\cite{regan}, \cite{sakamoto}.

Most of our sample galaxies with~$M_{mol}/M_{HI}>2$ do not show such
high-contrast and well-defined spiral structures as {\it M51}, but,
judging from the Digital Sky Survey images, a circumnuclear region
$1\div2$ $kpc$ in size can usually be distinguished by its high
brightness, which is indicative of intense star formation in the
dense central part of a disk. High pressures, and, consequently, a
"molecularization" of the gas in the central region of the galaxy
could be due to the high surface density of gas combined with the
high density of the central part of the stellar disk, whose
gravitational field compresses the gas into a thin layer, increasing
its volume density.

Unfortunately, data on the radial distribution of the gas are
available for only a few of the galaxies in our sample. The best
studied example is {\it M100}, for which a high-resolution radial
$I_{CO}$ profile was obtained with the BIMA \cite{regan} and
Nobeyama Observatory \cite{sakamoto} interferometers. Most of the
gas in this galaxy is in the molecular state, and is concentrated in
an inner region with a diameter of $2\div3$ $kpc$ (although the gas
also remains mostly molecular at larger distances from the center).

The radial motion of the gas toward the center could be due to both
dynamical friction acting on giant molecular clouds and to the
interaction of gas with a currently or previously existing bar
(recall that more than half of our galaxies have appreciable bars).

\section{CONCLUSIONS}
Galaxies whose gas is predominantly in molecular form
($M_{mol}/M_{HI}>2$) occur among disk galaxies of all morphological
types, although {\it Sc~---Sd} types are among these objects. These
galaxies differ only slightly from galaxies with lower relative mass
of molecular gas in terms of their luminosities, surface
brightnesses, rotational velocities, and dust contents per unit mass
of gas. At the same time, galaxies that are rich in molecular gas
show a tendency to have bars and active nuclei (to be LINERs),
providing indirect evidence for a concentration of gas toward the
nucleus.

The high $M_{mol}/M_{HI}$ ratios in the galaxy sample we have
considered is not due to a loss of $HI$ or the acquisition of $H_2$,
but instead is the result of the transformation of most of the $HI$
into~$H_2$, since the total gas mass, $M_{gas}$, remains normal and
consistent with the specific angular momentum of the disk. However,
the $M_{mol}$ values for some of the sample galaxies derived from
{\it CO}-line intensities are probably appreciably overestimated due
to the use of a conversion factor that appears to be overestimated.

With only a few exceptions, galaxies with high $M_{mol}/M_{HI}$
ratios are characterized by moderate star-formation rates and
comparatively high star-formation efficiencies, $SFR/M_{gas}$.
However, the mean star-formation rate per unit mass of molecular gas
in these galaxies is appreciably lower compared to galaxies with low
$M_{mol}/M_{HI}$. This means that, on average, the gas spends more
time in molecular form before it becomes atomic again.

As far as the integral properties of our sample galaxies are normal,
their high content of molecular gas must be due to specific
properties of the gas in local regions of the disk. The high
fraction of molecular gas appears to be due to two interrelated
factors. The first of these is a high density (pressure) of the gas
due to its concentration in the inner region of the disk (or in
spiral arms, if the galaxy possesses a high-contrast spiral
structure). In this case, the local dust density should also be
higher, facilitating the formation and preservation of molecules. We
think that there are no galaxies with a deficit of gas in their
central region among those with a predominance of molecular gas. The
second possible factor is the long duration of the molecular state
of the gas during the chain  $HI \to H_2 \to$~gravitationally
unstable clouds~$\to$~stars. Evidence for this is provided by the
systematically lower star-formation rate per unit mass of molecular
gas for the sample galaxies. If the gas is concentrated in the inner
regions of the galaxy, this delay of its transformation into stars
can be logically associated with the high angular velocity of
rotation in the inner disk region, which enhances the gravitational
stability of the gaseous disk against local perturbations, which, in
turn, leads to the accumulation of gas and its subsequent transition
into molecular form.

We note, however, that the concentration of molecular gas toward
the center of the galaxy is a fairly common (although according to
\cite{boker}, it is less pronounced for galaxies of the latest
morphological types). However, a predominance of molecular over
atomic gas is a rare phenomenon, implying that a central
concentration of a gas is essential, but no means a sufficient
condition for such a predominance.  To understand better the
causes for the transformation of  most of $HI$ into $H_2$ on a
galactic scale a more complete picture of gas evolution  remains
needed.

The work was supported by RFBR grant 04-02-16518.

\newpage
%
\begin{table}
\centering \caption{Galaxies with $M_{mol}/M_{HI}>2$.}
{\scriptsize \hspace{-10mm}
\begin{tabular}{cccccccccccccc}
 \hline \hline
\small n & Name & Type & Dist. & $R_o$    & L     & $D_{25}    $ & $\lg(M_{dust})  $ & $\lg(M_{HI})$   & $\lg(M_{mol})$   & $\lg(V_{rot})$ & $(U-B)_c$ & $(B-V)_c$ & Remarks \\
 & & & Mpc & Kpc & $L_{\odot}$ & Kpc & $M_{\odot}$ & $M_{\odot}$ & $M_{\odot}$ & km/s & & & \\

\hline \hline
$1$ & $NGC23     $ & $ 1.2$ & $  67$ & $0.57$ & $10.78$ & $1.56$ & $7.14$ & $9.95$ & $10.27$ & $2.45$ & --- & $0.72$ & B          \\
$2$ & $NGC142     $ & $ 3.0$ & $ 113$ & --- & $10.64$ & $1.53$ & $7.04$ & $9.73$ & $10.16$ & $2.10$ & --- & --- & B          \\
$3$ & $NGC695     $ & $-2.0$ & $ 140$ & --- & $11.15$ & $1.31$ & $7.64$ & $10.36$ & $10.81$ & $2.45$ & --- & --- &            \\
$4$ & $NGC828     $ & $ 1.0$ & $  78$ & --- & $10.88$ & $1.82$ & $7.69$ & $10.15$ & $10.75$ & $2.48$ & $0.32$ & $0.79$ &            \\
$5$ & $NGC972     $ & $ 2.0$ & $  23$ & $0.53$ & $10.35$ & $1.34$ & $6.95$ & $9.13$ & $9.75$ & $2.19$ & $0.07$ & $0.66$ &            \\
$6$ & $NGC1022    $ & $ 1.1$ & $  19$ & $0.26$ & $9.94$ & $1.16$ & $6.13$ & $8.53$ & $9.24$ & $2.02$ & $0.20$ & $0.69$ & B          \\
$7$ & $NGC1482    $ & $-0.9$ & $  24$ & --- & $9.72$ & $1.19$ & $6.58$ & $8.89$ & $9.68$ & $2.13$ & $-0.01$ & $0.86$ &            \\
$8$ & $UGC2982    $ & $ 5.6$ & $  75$ & --- & $10.75$ & $1.25$ & $7.28$ & $10.14$ & $10.47$ & $2.32$ & --- & --- & B          \\
$9$ & $IC2040     $ & $-1.0$ & $  16$ & --- & $9.06$ & $0.81$ & $5.04$ & $8.23$ & $8.73$ & $1.74$ & --- & --- &            \\
$10$ & $NGC1819    $ & $-2.0$ & $  63$ & --- & $10.49$ & $1.46$ & $6.91$ & $9.51$ & $10.13$ & $2.25$ & --- & --- & B          \\
$11$ & $NGC2146    $ & $ 2.3$ & $  16$ & --- & $10.42$ & $1.42$ & $6.80$ & $9.67$ & $10.06$ & $2.30$ & $0.18$ & $0.65$ & B          \\
$12$ & $NGC2775    $ & $ 1.6$ & $  19$ & --- & $10.46$ & $1.40$ & $6.89$ & $8.66$ & $9.49$ & $2.49$ & $0.32$ & $0.83$ &            \\
$13$ & $NGC3032    $ & $-1.9$ & $  24$ & $0.30$ & $9.66$ & $1.09$ & $5.79$ & $8.22$ & $8.91$ & $2.27$ & $0.09$ & $0.63$ & B          \\
$14$ & $NGC3147    $ & $ 3.9$ & $  44$ & --- & $11.02$ & $1.74$ & $7.51$ & $10.21$ & $10.60$ & $2.54$ & --- & $0.77$ &            \\
$15$ & $NGC3504    $ & $ 2.1$ & $  24$ & $0.40$ & $10.30$ & $1.26$ & $6.52$ & $8.78$ & $9.70$ & $2.25$ & $-0.03$ & $0.67$ & B          \\
$16$ & $NGC3683    $ & $ 5.0$ & $  28$ & $0.18$ & $10.17$ & $1.17$ & $6.76$ & $9.32$ & $9.66$ & $2.27$ & --- & --- & B          \\
$17$ & $NGC3758    $ & $ 2.7$ & $ 129$ & --- & $10.63$ & $1.31$ & $6.89$ & $9.47$ & $9.90$ & $2.43$ & --- & --- &            \\
$18$ & $NGC3860    $ & $ 2.0$ & $  81$ & --- & $10.48$ & $1.44$ & $7.01$ & $9.22$ & $9.54$ & $2.41$ & $0.25$ & $0.68$ &            \\
$19$ & $CGCG157-46   $ & $ 3.0$ & $  96$ & --- & $10.16$ & $1.38$ & $6.73$ & $9.03$ & $9.47$ & --- & --- & --- &            \\
$20$ & $NGC3953    $ & $ 4.0$ & $  18$ & $0.59$ & $10.63$ & $1.61$ & $7.26$ & $9.41$ & $9.96$ & $2.35$ & $0.12$ & $0.68$ & B          \\
$21$ & $NGC3993    $ & $ 3.1$ & $  71$ & --- & $10.54$ & $1.53$ & --- & $9.80$ & $10.61$ & $2.24$ & --- & --- &            \\
$22$ & $NGC4064    $ & $ 1.4$ & $  15$ & --- & $9.81$ & $1.24$ & $5.54$ & $7.96$ & $8.48$ & $1.90$ & $0.09$ & $0.67$ & B          \\
$23$ & $UGC07064    $ & $ 3.3$ & $ 109$ & --- & $10.62$ & $1.44$ & $7.63$ & $9.60$ & $10.21$ & $2.42$ & $-0.02$ & $0.60$ & B          \\
$24$ & $NGC4102    $ & $ 3.0$ & $  15$ & --- & $9.94$ & $1.14$ & $6.36$ & $8.75$ & $9.37$ & $2.23$ & --- & --- & B          \\
$25$ & $MCG5-29-45   $ & $ 3.2$ & $ 112$ & --- & $10.55$ & $1.42$ & $6.74$ & $9.87$ & $10.78$ & $1.97$ & $-0.51$ & $0.17$ & B, VV      \\
$26$ & $NGC4245    $ & $ 0.1$ & $  15$ & $0.30$ & $9.65$ & $1.12$ & $5.46$ & $7.80$ & $8.35$ & $2.12$ & $0.43$ & $0.85$ & B          \\
$27$ & $NGC4274    $ & $ 1.7$ & $  16$ & $0.63$ & $10.20$ & $1.52$ & $6.32$ & $8.96$ & $9.31$ & $2.37$ & $0.35$ & $0.83$ & B          \\
$28$ & $NGC4298    $ & $ 5.2$ & $  18$ & $0.38$ & $10.10$ & $1.18$ & $6.83$ & $8.99$ & $9.48$ & $2.10$ & --- & $0.61$ & V          \\
$29$ & $NGC4310    $ & $-1.0$ & $  15$ & --- & $9.05$ & $0.97$ & $5.35$ & $7.88$ & $8.31$ & $1.90$ & --- & --- & B          \\
$30$ & $NGC4314    $ & $ 1.0$ & $  16$ & $-0.06$ & $10.11$ & $1.27$ & $5.67$ & $7.14$ & $8.98$ & $2.33$ & $0.27$ & $0.81$ & B          \\
$31$ & $NGC4321    $ & $ 4.0$ & $  24$ & $0.95$ & $11.04$ & $1.72$ & $7.22$ & $9.93$ & $10.46$ & $2.28$ & $-0.05$ & $0.65$ & B, V       \\
$32$ & $NGC4394    $ & $ 2.9$ & $  15$ & $0.37$ & $9.91$ & $1.17$ & $5.94$ & $8.56$ & $8.96$ & $2.33$ & $0.19$ & $0.81$ & B, V       \\
$33$ & $NGC4414    $ & $ 5.1$ & $  13$ & $0.23$ & $10.14$ & $1.12$ & $6.50$ & $9.39$ & $9.70$ & $2.35$ & $0.10$ & $0.77$ & V          \\
$34$ & $NGC4418    $ & $ 1.0$ & $  31$ & --- & $9.69$ & $1.13$ & $6.31$ & $8.83$ & $9.34$ & $1.70$ & --- & --- &            \\
$35$ & $NGC4419    $ & $ 1.1$ & $  22$ & $0.13$ & $10.31$ & $1.33$ & $6.39$ & $8.41$ & $9.83$ & $2.05$ & $0.38$ & $0.81$ & B, V       \\
$36$ & $NGC4448    $ & $ 1.8$ & $  12$ & $0.04$ & $9.54$ & $0.94$ & $5.51$ & $7.81$ & $8.41$ & $2.32$ & $0.33$ & $0.86$ & B          \\
$37$ & $NGC4450    $ & $ 2.3$ & $  30$ & $0.81$ & $10.89$ & $1.64$ & $7.08$ & $9.05$ & $9.39$ & $2.30$ & --- & $0.76$ & V          \\
$38$ & $NGC4457    $ & $ 0.5$ & $  13$ & $0.49$ & $9.76$ & $1.03$ & $5.57$ & $8.32$ & $9.10$ & $2.04$ & $0.23$ & $0.81$ & B, V       \\
$39$ & $NGC4501    $ & $ 3.3$ & $  34$ & $0.89$ & $11.39$ & $1.83$ & $7.68$ & $10.04$ & $10.59$ & $2.46$ & $0.29$ & $0.63$ & V          \\
$40$ & $NGC4548    $ & $ 3.1$ & $   8$ & $0.07$ & $10.59$ & $1.10$ & $6.14$ & $8.29$ & $8.76$ & $2.29$ & $0.38$ & $0.75$ & B          \\
$41$ & $NGC4579    $ & $ 2.8$ & $  23$ & $0.68$ & $10.82$ & $1.57$ & $6.86$ & $9.06$ & $9.86$ & $2.44$ & $0.41$ & $0.75$ & B          \\
$42$ & $NGC4580    $ & $ 1.6$ & $  16$ & $0.15$ & $9.55$ & $0.96$ & $5.91$ & $7.66$ & $8.15$ & $2.02$ & --- & --- & B          \\
$43$ & $NGC4689    $ & $ 4.7$ & $  25$ & $0.66$ & $10.44$ & $1.51$ & $6.62$ & $9.09$ & $9.81$ & $2.09$ & --- & $0.60$ & V          \\
$44$ & $NGC4710    $ & $-0.8$ & $  20$ & $0.24$ & $10.13$ & $1.46$ & $6.17$ & $8.06$ & $9.10$ & $2.17$ & $0.29$ & $0.78$ &            \\
$45$ & $NGC4818    $ & $ 2.0$ & $  15$ & $0.41$ & $10.02$ & $1.20$ & $5.87$ & $7.98$ & $9.12$ & $2.12$ & $0.15$ & $0.76$ & B          \\
$46$ & $NGC4848    $ & $ 5.6$ & $ 104$ & --- & $10.83$ & $1.66$ & $6.96$ & $9.27$ & $9.58$ & $2.08$ & $-0.14$ & $0.52$ & B          \\
$47$ & $NGC4858    $ & $ 3.0$ & $ 137$ & --- & $10.30$ & $1.31$ & $6.91$ & $9.05$ & $9.70$ & --- & $-0.20$ & $0.36$ & B, C       \\
$48$ & $IC4040     $ & $ 7.2$ & $ 115$ & --- & $10.45$ & $1.44$ & $7.09$ & $9.07$ & $9.66$ & $1.73$ & $-0.20$ & $0.39$ & B, C       \\
$49$ & $NGC4922    $ & $-4.4$ & $ 104$ & --- & $10.73$ & $1.58$ & $6.86$ & $9.35$ & $10.03$ & --- & $0.54$ & $0.78$ & VV, C      \\
$50$ & $NGC4984    $ & $-0.8$ & $  17$ & $1.34$ & $9.86$ & $1.18$ & $5.75$ & $8.24$ & $9.22$ & $2.12$ & $0.28$ & $0.83$ & B          \\
$51$ & $IC860     $ & $ 2.4$ & $  58$ & --- & $10.03$ & $1.19$ & $6.76$ & $7.87$ & $9.03$ & $2.37$ & --- & --- &            \\
$52$ & $NGC5054    $ & $ 4.1$ & $  24$ & --- & $10.51$ & $1.53$ & $6.73$ & $9.58$ & $9.96$ & $2.27$ & $0.08$ & $0.63$ &            \\
$53$ & $UGC8399    $ & $ 3.0$ & $ 107$ & --- & $10.37$ & $1.43$ & $6.57$ & $9.74$ & $10.62$ & $1.76$ & --- & --- & B          \\
$54$ & $MCG5-32-20   $ & $ 8.0$ & $ 103$ & --- & $10.30$ & $1.25$ & $5.96$ & $9.51$ & $10.13$ & $2.26$ & $-0.40$ & $0.32$ & VV         \\
$55$ & $NGC5187    $ & $ 3.0$ & $ 105$ & --- & $10.36$ & $1.49$ & $6.97$ & $9.99$ & $10.57$ & $2.26$ & --- & --- &            \\
$56$ & $NGC5653    $ & $ 3.0$ & $  54$ & --- & $10.62$ & $1.43$ & $7.11$ & $9.69$ & $10.08$ & $2.38$ & $-0.11$ & $0.62$ &            \\
$57$ & $NGC5678    $ & $ 3.3$ & $  31$ & --- & $10.64$ & $1.48$ & $7.04$ & $9.53$ & $10.13$ & $2.30$ & --- & --- & B          \\
$58$ & $NGC5676    $ & $ 4.7$ & $  34$ & $0.65$ & $10.79$ & $1.58$ & $7.18$ & $9.94$ & $10.26$ & $2.39$ & $-0.01$ & $0.57$ &            \\
$59$ & $NGC5936    $ & $ 3.2$ & $  59$ & $0.57$ & $10.59$ & $1.36$ & $7.07$ & $9.63$ & $10.12$ & $2.46$ & $-0.12$ & $0.52$ & B          \\
$60$ & $NGC6000    $ & $ 4.0$ & $  30$ & --- & $10.31$ & $1.22$ & $6.95$ & $9.82$ & $10.13$ & $2.27$ & --- & --- & B          \\
$61$ & $NGC6240    $ & $ 0.0$ & $ 105$ & --- & $10.92$ & $1.80$ & $7.46$ & $10.01$ & $10.69$ & $2.30$ & $0.25$ & $0.69$ & VV         \\
$62$ & $NGC6574    $ & $ 3.9$ & $  35$ & --- & $10.55$ & $1.17$ & $6.86$ & $9.24$ & $10.09$ & $2.40$ & $0.02$ & $0.63$ & B          \\
$63$ & $NGC6951    $ & $ 3.9$ & $  24$ & $0.75$ & $10.89$ & $1.39$ & $6.92$ & $9.72$ & $10.11$ & $2.35$ & --- & $0.60$ & B          \\
$64$ & $NGC7225    $ & $-0.3$ & $  68$ & --- & $10.51$ & $1.59$ & $7.03$ & $9.38$ & $10.07$ & $2.21$ & --- & --- &            \\
$65$ & $NGC7770    $ & $ 0.0$ & $  62$ & --- & $10.17$ & $1.18$ & --- & $9.81$ & $10.49$ & $2.59$ & --- & $0.47$ &            \\
$66$ & $MCG3-60-36   $ & $ 1.0$ & $  79$ & --- & $10.28$ & $1.29$ & $7.17$ & $10.03$ & $10.62$ & $2.10$ & --- & --- &            \\
\hline  \label{tab1}
\end{tabular}
 }
\end{table}
\end{document}